\newif\ifsubmode
\submodefalse

\newif\ifprintfig
\printfigfalse

\ifsubmode
  \documentstyle[12pt,aasms4,epsf]{article}
  \received{}
  \accepted{}
  \journalid{}{}
  \articleid{}{}
\else
  \documentstyle[11pt,aaspp4,epsf]{article}
  \slugcomment{{\sl submitted to The Astrophysical Journal; Feb 17, 1997.}}
\fi

\lefthead{van der Marel, Sigurdsson \& Hernquist}
\righthead{Dynamical stability of N-body models for M32}

\def\etal{{et al.~}}
\def\lta{\lesssim}
\def\gta{\gtrsim}

\def\pc{\>{\rm pc}}
\def\Mpc{\>{\rm Mpc}}
\def\Msun{\>{\rm M_{\odot}}}

\def\Mdark{M_{\bullet}}

\def\cP{{\cal P}}
\def\cR{{\cal R}}
\def\fm{f_{\rm m}}

\def\refmassdens{1}

\def\refmassspec{4}

\begin{document}

\title{Dynamical stability of N-body models for M32\\
       with a central black hole}

\author{Roeland P.~van der Marel\altaffilmark{1}}
\affil{Institute for Advanced Study, Olden Lane, Princeton, NJ 08540}

\author{Steinn Sigurdsson\altaffilmark{2}}
\affil{Institute of Astronomy, Madingley Road, Cambridge CB3 0HA, 
       England}

\author{Lars Hernquist\altaffilmark{3}}
\affil{Department of Astronomy and Astrophysics,
       University of California, Santa Cruz, CA 95064}


\altaffiltext{1}{Hubble Fellow}
\altaffiltext{2}{Marie Curie Fellow}
\altaffiltext{3}{Presidential Faculty Fellow}


\ifsubmode\else
\clearpage\fi


\ifsubmode\else
\baselineskip=14pt
\fi


\begin{abstract}
We study the stability of stellar dynamical equilibrium models for
M32. Kinematic observations show that M32 has a central dark mass of
$\sim 3 \times 10^6 \Msun$, most likely a black hole, and a
phase-space distribution function that is close to the `two-integral'
form $f=f(E,L_z)$. M32 is also rapidly rotating; 85--90\% of the stars
have the same sense of rotation around the symmetry axis. Previous
work has shown that flattened, rapidly rotating two-integral models
can be bar-unstable. We have performed N-body simulations to test
whether this is the case for M32.  This is the first stability
analysis of two-integral models that have both a central density cusp
and a nuclear black hole.

Particle realizations with $N=\>$512,000 were generated from
distribution functions that fit the photometric and kinematic data of
M32. We constructed equal-mass particle realizations, and also
realizations with a mass spectrum to improve the central resolution.
Models were studied for two representative inclinations,
$i=90^{\circ}$ (edge-on) and $i=55^{\circ}$, corresponding to
intrinsic axial ratios of $q=0.73$ and $q=0.55$, respectively.  The
time evolution of the models was calculated with a `self-consistent
field' code on a Cray T3D parallel supercomputer. We find both models
to be dynamically stable. This implies that they provide a physically
meaningful description of M32, and that the inclination of M32 (and
hence its intrinsic flattening) cannot be strongly constrained through
stability arguments.

Previous work on the stability of $f(E,L_z)$ models has shown that the
bar-mode is the only possibly unstable mode for systems rounder than
$q \approx 0.3$ (i.e., E7), and that the likelihood for this mode to
be unstable increases with flattening and rotation rate. The
$f(E,L_z)$ models studied for M32 are stable, and M32 has a higher
rotation rate than nearly all other elliptical galaxies. This suggests
that $f(E,L_z)$ models constructed to fit data for real elliptical
galaxies will generally be stable, at least for systems rounder than
$q \gta 0.55$, and possibly for flatter systems as well.
\end{abstract}


\keywords{black hole physics ---
          galaxies: elliptical and lenticular, cD ---
          galaxies: individual (M32) ---
          galaxies: kinematics and dynamics ---
          galaxies: nuclei ---
          galaxies: structure.}

\clearpage


\section{Introduction}

Active galaxies and quasars are thought to be powered by the presence
of massive black holes (BHs) in their nuclei (e.g., Rees 1997). Such
BHs are believed to exist in a large fraction of quiescent galaxies as
well (e.g., Tremaine 1996). The best dynamical BH detections in active
galaxies come from studies of gas kinematics; in quiescent galaxies
only stars are generally available as a tracer of the gravitational
potential. Ground-based stellar kinematical studies have yielded
tentative detections for BHs in a handful of nearby quiescent galaxies
(Kormendy \& Richstone 1995), and this evidence is now being improved
with the superior spatial resolution of the Hubble Space Telescope
(HST; Kormendy \etal 1996a,b; van der Marel \etal 1997a). However, the
interpretation of the observed stellar motions is never
straightforward; only projected quantities are observed and the
three-dimensional orbital structure is unknown. Construction of
detailed dynamical models is therefore required. Axisymmetric models
currently provide the state-of-the-art for this application. In such
models, the phase-space distribution function (DF) generally depends
on three integrals of motion (e.g., Binney \& Tremaine 1987), the two
classical integrals, $E$ and $L_z$ (the binding energy and the angular
momentum component along the symmetry axis, both per unit mass), and
the third integral, $I_3$. The latter cannot generally be expressed
analytically in terms of the phase-space coordinates, which makes the
construction of models with `three-integral' DFs, $f= f(E,L_z,I_3)$,
complicated and time-consuming, although not impossible (Dehnen \&
Gerhard 1993; Cretton \etal 1997; Gebhardt \etal 1997). As a result,
significant attention has been given in recent years to a special
class of axisymmetric models, those in which the DF depends only on
the two classical integrals of motion, $f=f(E,L_z)$. The attraction of
these `two-integral' models lies in the fact that their properties are
tractable semi-analytically (e.g., Hunter \& Qian 1993; Dehnen \&
Gerhard 1994). These models can be viewed as the axisymmetric
generalization of spherical isotropic models: the velocity dispersions
in the meridional plane are isotropic, i.e., $\sigma_r =
\sigma_{\theta}$, and the flattening of the models (when oblate) is
due to an excess of azimuthal motion.

The present paper deals with the nearby E3 galaxy M32, which has been
a strong candidate for containing a BH for many years. The highest
spatial resolution ground-based data are well fit by an $f(E,L_z)$
model with a 2--$3 \times 10^6 \Msun$ BH (van der Marel
\etal 1994; Qian \etal 1995; Dehnen 1995; Bender, Kormendy \& Dehnen 1996). 
Such a model fits the observed line-of-sight velocity profile shapes
without any adjustable parameters, suggesting that the DF of M32 is
indeed close to the $f(E,L_z)$ form. We recently completed the most
detailed study to date of M32 (van der Marel \etal 1997a), using new
HST data (van der Marel, de Zeeuw \& Rix 1997b) and fully general
three-integral modeling (van der Marel \etal 1997c; hereafter
vdM97c). The main results of these studies are that: (i) M32 {\it
must} have a central dark mass in the range $\Mdark = (3.4 \pm 1.6)
\times 10^6 \Msun$ (at the formal $99.73$\% confidence level),
enclosed within $\sim 0.3 \pc$, most likely a BH; and (ii) the
best-fitting models have a dynamical structure that is indeed similar
(although not identical) to that of an $f(E,L_z)$ model.

Although dynamical models with $f(E,L_z)$ DFs can reproduce most
features of the M32 data, they must also be stable if they are to be
physically meaningful. Only limited research has been done on the
stability of $f(E,L_z)$ models. Bending modes and lopsided modes in
extremely flattened systems (axial ratio $q \lta 0.2$) were discussed
by Merritt \& Fridman (1995). Kuijken \& Dubinski (1994) performed
N-body simulations of models with a King-type mass density profile and
an axial ratio varying between $q \approx 0.6$ in the center and $q
\approx 0.8$ in the outer parts. Models with little or no mean
streaming were stable, but rapidly rotating models were found to be
unstable to the formation of a bar. Preliminary results on the
bar-stability in $f(E,L_z)$ models with central density cusps, but
without central dark objects, were presented by Dehnen (1996). His
N-body simulations indicate that: (i) non-rotating models are stable,
even if they are as flat as $q \approx 0.3$; and (ii) maximally
rotating models tend to be bar-unstable if they are flatter than $q
\approx 0.5$. The most detailed study to date of the stability of
$f(E,L_z)$ models has been that for a family of Kuzmin-Kutuzov models
by Sellwood \& Valluri (1997). They identify all the most important
unstable modes as a function of shape and rotational support. Their
results confirm the main conclusions on bar modes found by earlier
authors, and they also present detailed studies of lopsided, bending
and other unstable modes. The introduction of their paper provides a
useful reference to other recent work on the stability of hot stellar
systems (see also the introductory chapter of Robijn 1995). None of
the previous studies included both a central density cusp and a
central black hole, as is appropriate for M32.

If M32 is being viewed edge-on, its intrinsic axial ratio is $q
\approx 0.73$. The galaxy must be intrinsically flatter if it is
inclined with respect to the line of sight. The observed rotation
velocities imply that, over the radial range $\lta 20''$ for which
kinematical data is available, 85--90\% of the stars (more-or-less
independent of the assumed inclination) must have the same sense of
rotation around the symmetry axis (van der Marel \etal 1994). The
previous work on the stability of $f(E,L_z)$ models suggests that the
M32 models should certainly be stable to, e.g., lopsided and bending
modes (unless for extreme inclinations), but that they might be
unstable to bar modes.  We therefore tested the stability of the
$f(E,L_z)$ models for M32 through N-body simulations on a parallel
supercomputer. Our results are described in what follows.  Section~2
summarizes our choice of models, Sections 3 describes the N-body
technique, and Section~4 discusses the results of the
simulations. Conclusions are presented in Section~5. An appendix
outlines the construction of the initial conditions.

\section{Models}

\subsection{Distribution functions}

The dynamical models that we study in this paper are based on those
presented in vdM97c. The {\it luminous} mass density is assumed to be
axisymmetric and of the form
\begin{equation}
  \rho(R,z) = \rho_0 \> (m/b)^{\alpha} \> [1+(m/b)^2]^{\beta} 
                                       \> [1+(m/c)^2]^{\gamma} , \qquad
   m^2 \equiv R^2 + (z/q)^2 .
\label{massdens}
\end{equation}
The intrinsic axial ratio $q$ is related to the projected axial ratio
$q_p$ and the inclination $i$ according to $q_p^2 = \cos^2 i + q^2
\sin^2 i$. The parameters are set to $\alpha=-1.435$, $\beta = -0.423$, 
$\gamma = -1.298$, $b=0.55''$, $c=102.0''$, $q_p=0.73$, $\rho_0 = j_0
\Upsilon M_{\odot} / L_{\odot,V}$, $j_0 = 0.463 \times 10^5 (q_p/q)
L_{\odot,V} \pc^{-3}$, and the distance to M32 is assumed to be $0.7
\Mpc$. The quantity $j$ is the luminosity density, and $\Upsilon$ is the 
average V-band mass-to-light ratio of the stellar population. Both
$\Upsilon$ and $q_p$ are assumed to be constant with radius. This
model provides an accurate fit to the observed surface brightness
profile of M32, including the available HST photometry in the central
few arcsec (Lauer \etal 1992). The adopted constant $q_p = 0.73$
provides a good approximation in the central $\sim 10''$, although at
larger radii the observed ellipticity increases slowly to $0.86$ at
$\sim 100''$. The model has finite mass, and two free parameters: the
inclination $i$ and the mass-to-light ratio $\Upsilon$. The results of
this paper do not depend on the particular
parametrization~(\ref{massdens}) that is used to characterize the M32
luminosity density.

The (positive) gravitational potential of M32 is assumed to be $\Psi =
\Psi_{\rm lum} + \Psi_{\rm dark}$, where $\Psi_{\rm lum}$ is the
potential generated by the luminous matter with mass
density~(\ref{massdens}), and $\Psi_{\rm dark}$ is the potential of a
nuclear massive dark object. The latter is taken to be the `softened
point mass' potential
\begin{equation}
   \Psi_{\rm dark} = G \Mdark (r^2 + \epsilon^2)^{-1/2} .
\label{potent}
\end{equation}
This is the potential generated by a cluster with a Plummer model mass
density (Binney \& Tremaine 1987). For the case of a nuclear BH (i.e.,
a point mass) the potential is Keplerian and $\epsilon = 0$. We do not
include the potential of a dark halo around M32; there are no (strong)
observational constraints on the possible presence and characteristics
of such a dark halo. Also, if anything, adding a dark halo probably
has a stabilizing effect (Ostriker \& Peebles 1973; Stiavelli \&
Sparke 1991), and omitting a dark halo is therefore conservative in
the present context.

In vdM97c, models with two- and three-integral DFs are constructed to
interpret the available kinematic M32 data. Two-integral models do not
reproduce all features of the data, but come close. The more general
three-integral modeling shows indeed that the best-fitting models do
not have two-integral DFs. However, they have a similar dynamical
structure: $\overline{v_{\phi}^2} > \overline{v_{\theta}^2} \gta
\overline{v_{r}^2}$, whereas for two-integral models
$\overline{v_{\phi}^2} > \overline{v_{\theta}^2} =
\overline{v_{r}^2}$. Because the construction of $N$-body initial conditions 
is significantly more complicated for three-integral than for
two-integral DFs, we restrict ourselves here to testing only the
stability of the two-integral models for M32. We do not expect the
results to be much different for more general models.

For a given gravitational potential there is a unique {\it even} DF
$f_{\rm e}(E,L_z)$ that reproduces the mass density $\rho(R,z)$. For
our models we calculated this DF as described in vdM97c. The {\it
total} DF is the sum of the part $f_{\rm e}$ that is even in $L_z$ and
the part $f_{\rm o}$ that is odd in $L_z$. In principle $f_{\rm o}$ is
determined completely by the mean streaming velocities
$\overline{v_{\phi}}(R,z)$, but these are not constrained well enough
by the data to make an inversion feasible. We therefore adopt a
convenient parametrization for the odd part:
\begin{equation}
  f_{\rm o}(E,L_z) = (2\eta - 1 ) \> f_{\rm e}(E,L_z) \> 
         h_a [L_z/L_{\rm max}(E)] ,
\label{fodddef}
\end{equation}
where $L_{\rm max}(E)$ is the maximum angular momentum at a given
energy. The function $h_a$ is a smoothed step function as defined in
van der Marel \etal (1994), with the parameter $a$ regulating the
step-smoothness. For any given potential, the free parameters $\eta$
and $a$ can be specified so as to best fit the observed rotation
velocities. Our results should not depend sensitively on the
particular choice of parametrization adopted here.

\subsection{Initial Conditions}

We chose to simulate models with $\Mdark = 3 \times 10^6 \Msun$ and
$\epsilon=0.04''$ ($=0.136 \pc$), using $N=\>$512,000 particles. The
use of a softened dark object in this context (rather than a point
mass) was motivated by numerical considerations. However, the adopted
value of $\epsilon$ is not inconsistent with the kinematic
observations (vdM97c; the size of the smallest HST/FOS aperture is
$0.086''$). There are no independent constraints on the inclination of
M32, and we therefore studied models with two representative values:
$i=90^{\circ}$ (edge-on) and $i=55^{\circ}$. These models have
intrinsic axial ratios $q=0.73$ and $q=0.55$, respectively. For both
inclinations we characterized the odd part of the DF by $\eta=1$ and
$a=5.5$, which provides an adequate fit to the observed rotation
velocities. The mass-to-light ratio was chosen to be $\Upsilon = 2.51$
for $i=90^{\circ}$, and $\Upsilon = 2.55$ for $i=55^{\circ}$. A more
complete study of the influence of the assumed inclination on the
stability of the models would be interesting, but also expensive in
terms of computing resources.

\placefigure{f:histnum}

Appendix~A describes how initial conditions were generated from the
DF. The most straightforward approach is to use equal-mass
particles. However, even with $N=\>$512,000 equal-mass particles, the
resolution near the center is poor. The solid histogram in Figure~1
shows the number of particles as function of radius $r$ for the
$i=55^{\circ}$ model with equal-mass particles. To improve the
resolution it is useful to adopt a spectrum of particle masses, such
that particles near the center are on average less massive. This can
be achieved by choosing the particle mass $\mu$ to be a decreasing
function of the binding energy $E$. A convenient choice is
\begin{equation}
  \mu(E) \propto \lbrace \> [R_{\Psi}(E)+r_0] \> / \> [R_{\Psi}(E)+r_1] 
                         \> \rbrace^{\lambda} ,
\label{massspec}
\end{equation}
where $\Psi(R,z)$ is the gravitational potential, and $R_{\Psi}(E)$ is
defined by $\Psi(R_{\Psi}(E),0) = E$. After some experimentation, a
mass spectrum was adopted with parameters $r_0=0.01''$, $r_1=100''$
and $\lambda=1$. This yields a particle distribution that well samples
the density cusp around the dark nuclear object, as well as the main
body of the galaxy (Figure~1, dotted curve). For these multi-mass
models (with $N=\>$512,000), particles with $r \ll r_0$ have $\mu
\approx 5\Msun$, while particles with $r \gg r_1$ have $\mu \approx 5
\times 10^4 \Msun$. Figure~2 shows a scatter plot of the particle mass
$\mu$ as a function of radius for the $i=55^{\circ}$ multi-mass model.

\placefigure{f:scatter}

\section{N-body method}

We performed N-body simulations on the Cray T3D parallel supercomputer
at the Pittsburgh Supercomputing Center. The `self-consistent field'
(SCF) code developed by Hernquist \& Ostriker (1992) was used,
implemented for parallel architectures as described in Hernquist,
Sigurdsson \& Bryan (1995) and Sigurdsson \etal (1997).

To improve the accuracy of the particle integration near the black
hole, and to allow the smoothing length for the black hole to be
small, the previously used standard time centered leap-frog integrator
was replaced by a fourth order Hermite integration scheme (Makino
1991; Makino \& Aarseth 1992).  While the leapfrog has the virtue of
being both simple and symplectic, a fixed time step leapfrog is
vulnerable to spurious numerical scattering of orbits when potential
gradients are large, e.g., in steep density cusps and deep in
Keplerian potentials. Choosing a small enough fixed time step for the
leapfrog integrator to be completely free of numerical scattering in
the center is highly inefficient. A Hermite scheme provides a higher
order, but relatively cheap integration scheme which excels at
adapting to rapidly changing force terms. Since the Hermite scheme is
not symplectic, global energy conservation is limited by the Poisson
noise in the global potential due to the discrete realization of the
potential field (see Sigurdsson \& Mihos 1997 for a discussion).

A variable time step scheme was used, where the minimum time step was
set by the minimum time step for any particle, using Aarseth's
condition on the force derivatives (Aarseth 1985; Makino \& Aarseth
1992). To improve accuracy near the center (where time scales are
short), while not slowing down the entire calculation too much, a
modified two level `multistep' approach was used (Sigurdsson \& Mihos
1997) in which the accelerations due to the central massive object and
the self-gravity of the innermost group of particles are updated more
often than for the bulk of the particles at large radii. This approach
speeds up the calculations by an order of magnitude relative to a
single step integration. Compared to the multistep approach that we
used previously (Hernquist, Sigurdsson \& Bryan 1995) this approach is
about twice as fast at fixed integration precision. The adaptive time
step scheme is essential when the cusp is well resolved by the
particle realization and the black hole smoothing length is small, to
follow the particles near the center where forces and jerks are large.

The expansion uses a Hernquist (1990) model as its lowest order term,
for which $\rho \propto r^{-1}$ at small radii. For the case of M32
the scale-length of this model was chosen to be $a=8.73''$. This
minimizes the contribution to the potential from the higher-order
terms in the expansion. The calculations are done with units in which
$G=M=a=1$. In these dimensionless units, the smoothing length is
$\epsilon = 4.6\times 10^{-3}$, and the central dark mass is $\Mdark =
7.2 \times 10^{-3}$. In this system of units, time is measured in
units of $1.2 \times 10^6$ yrs, and the periods of circular orbits at
radii of $0.1''$, $1''$ and $10''$ are $0.095$, $1.4$ and $18$,
respectively.

We performed small simulations with 25,600 and 51,200 particles to
test the dependence of the results on the number of radial and angular
terms in the SCF expansion. Based on these results we adopted $n_{\rm
max} = 8$ and $l_{\rm max}=8$ for the final simulations. Non-zero $m$
terms were included in the expansion to allow any non-axisymmetric
instabilities to develop. Figure~3 illustrates the accuracy of the SCF
expansion, by comparing its circular velocity to that of the true
model potential. At small radii the {\it stellar} contribution to the
circular velocity is poorly represented, due to the finite number of
radial terms in the SCF expansion; in fact, increasing $n_{\rm max}$
by a small amount does not lead to a significantly improved
fit. However, the {\it total} circular velocity is well represented
for all $r \gta \epsilon$, because it is dominated by the central dark
object at the small radii where the SCF expansion fails. Only at radii
$r \ll \epsilon$ does the numerical approximation to the total
circular velocity become poor, and only as a result of the numerically
imposed softening (which causes the circular velocity at $r \ll
\epsilon$ to be dominated by the stellar density cusp, rather than by
the dark object).

\placefigure{f:vcirc}

In principle, the potential could have been represented more accurately
by using another set of basis functions. A basis set that has $\rho
\propto r^{\alpha}$ at small radii for its lowest order term, with
$\alpha=-1.435$ as in equation~(\ref{massdens}), would probably
provide a better description for M32. Such basis sets can indeed be
constructed (Saha 1993; Dehnen 1996; Zhao 1996), but in practice are
generally more difficult to implement and slower in use. The fact that
in our application the potential at small radii is dominated by the
dark object, justifies our use of the more conventional Hernquist
\& Ostriker (1992) basis set.

We also used smaller simulations to make a proper choice for the time
integration parameters. In the final simulations the smallest step
size (near the center) was typically $\sim 4 \times 10^{-4}$ in
dimensionless units, with the step size at larger radii being $\sim
40$--50 times larger. Global energy conservation was about 0.5\% over
$\Delta t = 100$.

One of the more useful diagnostics to monitor in the simulations is
the evolution of the axial ratios of the mass distribution as a
function of radius. For this we used an iterative scheme similar to
that used by Dubinski \& Carlberg (1991). A mass bin is adopted that
contains a fixed fraction of the total mass (e.g., the central
0--10\%). The moment of inertia tensor is calculated for the particles
that fall in the bin, and diagonalization then yields the axial ratios
$q_1$ and $q_2$ (the eigenvalues of the tensor) and the angles of the
principal axes with respect to the initial coordinate axes. Iteration
is required because the sorting of the particles in the ellipsoidal
radial coordinate depends on the angles and axial ratios. We found
that the algorithm has convergence problems when the number of
particles is small, or at small radii. To improve this, we made
modifications to the basic algorithm to allow the mass fraction to
vary, to ensure that the centroid of the particles in the bin was at
all times well inside their mean ellipsoidal radius. Iteration was
done by taking the geometric mean of the derived axial ratio and the
previously derived axial ratio. This provides for stabler convergence
of the shape parameters, and allowed sensible axial ratio estimates to
be obtained for a large range in radius (see also Sigurdsson \& Mihos
1997).

\section{Results}

We ran an equal-mass model and a multi-mass model for each
inclination. The equal-mass simulations were run to $t=100$ for the
$i=90^{\circ}$ model, and to $t=50$ for the $i=55^{\circ}$ model, to
explore global stability and stability at large radii. The multi-mass
models were run to $t=20$ to explore stability at small radii. The
main result is that {\it the models for both inclinations are
dynamically stable.}  We illustrate this by discussing and displaying
the time evolution for the $i=55^{\circ}$ model. The results for the
edge-on model are similar.

\placefigure{f:dens}
\placefigure{f:axialrat}

Figure~4 shows the initial and final mass densities. Figure~5 shows
the evolution of the axial ratios $q_1$ and $q_2$ for four radial
bins. At small radii we display the results from the multi-mass
models, and at large radii those from the equal-mass models. There are
no evolutionary trends in the mass distribution. The axial ratios are
constant at their values for the continuous limit ($q_1 = 1$, $q_2 = q
= 0.55$), to within the accuracy imposed by the finite number of
particles. The orientation of the principal axes of the mass
distribution also showed no trends with time.

\placefigure{f:Anlm}

A more sensitive diagnostic of possible instabilities is provided by
an analysis of the individual expansion terms of the mass
distribution. Such an analysis is straightforward, because the SCF
technique makes a harmonic decomposition of the density and potential
at each time step. Figure~6 shows the time evolution of some of the
expansion terms, in particular those that could have been expected to
show signatures of an instability. However, there is no sign of even a
weak lopsided ($m=1$) or bar ($m=2$) instability. Other expansion
terms showed the same result of constant amplitudes with time, to
within the noise limits imposed by the discreteness of the model.

\placefigure{f:kinematics}

Not only the mass distribution is stable, but also the phase-space
structure of the model. Figure~7 shows for the multi-mass model the
initial and final mean azimuthal velocity $\overline{v_{\phi}}$, and
the RMS velocities $[\overline{v_{\phi}^2}]^{1/2}$,
$[\overline{v_{\theta}^2}]^{1/2}$ and $[\overline{v_{r}^2}]^{1/2}$,
all as a function of the spheroidal radius $m$. There is some
evolution in the velocities at radii close to the softening scale
$\epsilon$, due to small number statistics and the finite resolution
of the SCF expansion. Overall, however, the velocity structure of the
model is stable. The equality of $[\overline{v_{\theta}^2}]^{1/2}$ and
$[\overline{v_{r}^2}]^{1/2}$ is consistent with the $f(E,L_z)$ form of
the DF. Figure~8 shows that the model remains in virial equilibrium
throughout, as expected from the stability of both the mass
distribution and the kinematic structure.

\placefigure{f:virial}

\section{Conclusions}

Two integral $f(E,L_z)$ models with a central BH provide a reasonable
fit to the observed kinematics of M32. Previous work indicated that
$f(E,L_z)$ models with as much rotation as M32 could be bar-unstable,
but none of the previous studies took into account both a central cusp
in the mass density and a central BH. We therefore constructed
particle realizations of $f(E,L_z)$ models for M32, and calculated
their time evolution using a SCF N-body code on a Cray T3D parallel
supercomputer. Models were studied for two representative
inclinations, $i=90^{\circ}$ and $i=55^{\circ}$, corresponding to
intrinsic axial ratios of $q=0.73$ and $q=0.55$, respectively. Both
these models were found to be dynamically stable.

This result has two implications. First, the equilibrium models that
best fit the kinematic data for M32 are stable, and therefore
physically meaningful. Second, imposing the requirement of stability
on the models does not put very stringent constraints on the
inclination or the intrinsic axial ratio of the models.

Previous work on the stability of $f(E,L_z)$ models has shown that the
$m=2$ bar-mode is the only mode that might be unstable for systems
rounder than $q \approx 0.3$, and that the susceptibility to this
instability increases with flattening and rotation rate (Dehnen 1996;
Sellwood \& Valluri 1997). Here we have shown that $f(E,L_z)$ models
for M32 as flat as $q=0.55$ are stable. Combined with the fact that
M32 has a higher rotation rate $V/\sigma$ than nearly all other
elliptical galaxies (M32 rotates more rapidly than an oblate isotropic
rotator; van der Marel \etal 1994), this suggests that all $f(E,L_z)$
models with $q \gta 0.55$ that are constructed to fit real galaxies
will generally be stable. This might in fact be true even for flatter
models, but that is not something that can be addressed with the
results presented here.

Nonetheless, the general applicability of $f(E,L_z)$ models to real
galaxies remains unclear. Two independent lines of observational
evidence suggest that elliptical galaxies, especially those flatter
than E2, cannot have $f(E,L_z)$ DFs. The models tend to possess too
much motion on the major axis relative to the minor axis (Binney,
Davies \& Illingworth 1990; van der Marel 1991), and line-of-sight
velocity profiles that are too flat-topped (Bender, Saglia \& Gerhard
1994). Either way, the relative simplicity of these models and their
demonstrated use for the case of M32, are likely to leave them as a
popular tool for the interpretation of stellar kinematic data for some
time to come.

A more detailed study of the general stability properties of
$f(E,L_z)$ models is outside the scope of the present paper. However,
it will certainly be of interest to study the influence of a central
dark object on the stability of flattened rotating models, and, more
generally, to search for evolution in the region close to the dark
object as it grows with time. Parallel N-body integration algorithms
such as those used here are well suited to address these issues.

The longest simulation we ran covered $1.2 \times 10^8$ yrs. This is
many dynamical times in the central region of M32, and therefore
sufficient to test dynamical stability. However, it is only 1\% of the
Hubble time, and therefore insufficient for a study of very slow
secular evolution. Although collisionless secular evolution need not
necessarily be present, M32 will certainly suffer some secular
evolution as a result of two-body relaxation; the relaxation time in
the inner regions is only $4 \times 10^9$ yrs, independent of radius
(Lauer \etal 1992). Simulations that address the effects of this
relaxation using a hybrid direct/SCF N-body code are in progress
(Hemsendorf, Spurzem \& Sigurdsson 1997). More complicated simulations
are required to also include the effects of mass loss due to stellar
evolution and stellar collisions, which operate on similar time
scales.


\acknowledgments

We are grateful to Tim de Zeeuw, Kathryn Johnston, Chris Mihos, Gerry
Quinlan, Hans-Walter Rix and Rainer Spurzem, for helpful discussions,
useful comments, and for collaboration in various related
projects. The calculations reported here were performed at the
Pittsburgh Supercomputing Center; computing resources were also
provided by the National Center for Supercomputing
Applications. Financial support was provided by: (a) NASA, through
grant number \#GO-05847.01-94A, and through a Hubble Fellowship
\#HF-1065.01-94A awarded to RPvdM, both from the Space Telescope
Science Institute which is operated by the Association of Universities
for Research in Astronomy, Incorporated, under NASA contract
NAS5-26555; (b) NSF, through grant ASC 93-18185; (c) the European
Union, through a TMR Cat.~30 Marie Curie Fellowship awarded to SS; (d)
the Presidential Faculty Fellows Program; (e) the Aspen Center for
Physics (which also provided hospitality).


\clearpage

\appendix

\section{Drawing initial conditions from the distribution function}

The N-body simulations require the generation of initial conditions
from known $f(E,L_z)$ DFs. This Appendix describes the techniques that
were used for this. The cases of equal-mass particles and of particles
with a mass spectrum depending on energy are described separately. The
former is more straightforward, while the latter can be useful to
increase the numerical resolution of a simulation near the center.

In the following, $\cP$ denotes a normalized probability distribution,
and $\cR_i $ are random numbers in the interval $[0,1]$.

\subsection{Equal-mass particles}

For equal-mass particles one can first draw the particle positions.
The ellipsoidal radius $m$ and ellipsoidal angle $\tau$ in the
meridional $(R,z)$ plane are defined through
\begin{equation}
  R = m \sin \tau , \qquad z = q m \cos \tau ,
\label{mtau}
\end{equation}
where $q$ is the axial ratio of the system. The total mass of the system
is 
\begin{equation}
   M = q \int_{0}^{2\pi} d\phi  
         \int_{0}^{\pi} \sin \tau \> d\tau 
         \int_{0}^{\infty} m^2 \rho(m) \> dm  ,
\label{totmass}
\end{equation}
where $\phi$ is the azimuthal angle and $\rho(m)$ is the mass density
defined in equation~(\ref{massdens}). The probability distributions of
$\phi$, $\tau$ and $m$ for a random mass element are therefore
\begin{eqnarray}
  &&\cP(\phi) = {1\over{2\pi}}    ,\quad (\phi \in [0,2\pi]) ;\qquad
  \cP(\tau)   = {1\over2}\sin\tau ,\quad (\tau \in [0,\pi]) ;\qquad 
     \nonumber \\
  &&\cP(m) = 4 \pi q \> m^2 \rho(m) \> / \> M , \quad (m \in [0,\infty)) .
\end{eqnarray}
The total mass $M$ is easily calculated numerically. The `rejection
method' (e.g., Press \etal 1992) with comparison function $F_{\rm
comp}(m) \equiv (a+bm)^{-3/2}$ can be used to draw a random $m$ from
$\cP(m)$. The constants $a$ and $b$ must be chosen such that $F_{\rm
comp}(m) \ge \cP(m)$ for all $m\ge 0$. A random position from the mass
density is then obtained as follows:
\begin{enumerate}
\item Draw a random number $\cR_1$. Set $m=(a/b) \> [(1-\cR_1)^{-2}-1]$.
\item Draw a random number $\cR_2$. Reject the previous step and return to 
step~1 if $\cR_2 > \cP(m) / F_{\rm comp}(m)$.
\item Draw random numbers $\cR_3$ and $\cR_4$. Set $\phi = 2 \pi \cR_3$ and
$\tau = \arccos \> (1-2\cR_4)$.
\end{enumerate}

After the particle positions have been drawn from the mass
density, the particle velocities must be drawn from the DF. The
discussion can be restricted to $f(E,L_z)$ models in which all stars
have $L_z \geq 0$. Such a DF, $\fm (E,L_z)$, is characterized by the
choice of a maximally rotating odd part. Once initial conditions for
this DF are available, initial conditions can be obtained for any
other DF $f(E,L_z)$ with the same even part (and arbitrary odd part)
by simply reversing the velocity of each particle with probability
$1-(f/\fm)$.

The maximum angular momentum at a given energy, $L_{\rm max}(E)$, is
the angular momentum of a circular orbit in the equatorial plane.
Define $\eta \equiv L_z / L_{\rm max} (E)$. The maximum $\eta$ for a
star with known position and energy is
\begin{equation}
  \eta_{\rm max} (R,z,E) = R \> [2(\Psi-E)]^{1/2} / L_{\rm max}(E) ,
\label{etamax}
\end{equation}
where $\Psi(R,z)$ is the (positive) gravitational potential. The
mass-density is the integral of the DF over velocity space, which can
be written as (e.g., Binney \& Tremaine 1987)
\begin{equation}
  \rho = {{2\pi}\over{R}} \int_{0}^{\Psi} dE 
         \int_{0}^{\eta_{\rm max}} \fm(E,\eta) \> L_{\rm max}(E) \> d\eta .
\label{rhoasint}
\end{equation}
The joint probability distribution of $(E,\eta)$ at a fixed position 
is therefore
\begin{equation}
  \cP(E,\eta) = {{2\pi}\over{R}} \fm(E,\eta) \> L_{\rm max}(E) \> / \> \rho
     , \qquad (E \in [0,\Psi],\quad \eta \in [0,\eta_{\rm max}]) .
\end{equation}
For flattened models, the DF is a monotonically increasing function of
$\eta$ (e.g., Qian \etal 1995). Hence, one may use the rejection
method with comparison function
\begin{equation}
  F_{\rm comp} = \max_{E \in [0,\Psi]} \cP(E,\eta_{\rm max}) ,
\end{equation}
which can be shown to be finite. It is convenient to calculate $F_{\rm
comp}$ on a grid in the meridional plane, such that its value at any
position can be obtained quickly through spline interpolation. A
random velocity vector for a particle with known position is then
obtained as follows:
\begin{enumerate}
\item Draw random numbers $\cR_1$ and $\cR_2$. Set $E= \cR_1 \Psi$ and
$\eta = \cR_2$.
\item Calculate $\eta_{\rm max}$ from equation~(\ref{etamax}). 
Draw a random number $\cR_3$. Reject the previous step and return to
step~1 if either $\eta > \eta_{\rm max}$ or $\cR_3 > \cP(E,\eta) /
F_{\rm comp}$.
\item Draw a random number $\cR_4$. Set $\xi = 2 \pi \cR_4$. Calculate 
the particle velocities from
\begin{equation}
  v_{\phi} = \eta L_{\rm max}(E) / R , \qquad
  v_{\rm m} = \sqrt{2(\Psi-E)-v_{\phi}^2} , \qquad
  v_r = v_{\rm m}\cos\xi , \qquad v_{\theta} = v_{\rm m}\sin\xi .
\label{velocities}
\end{equation}
\end{enumerate}

\subsection{Particles with a mass spectrum}

Particle positions and velocities must be drawn simultaneously if the
particle mass $\mu$ depends on the phase space coordinates,
$\mu=\mu({\vec r},{\vec v})$. The phase space number density is the
ratio of the DF and the particle mass. With
equations~(\ref{totmass}), (\ref{rhoasint})
and~(\ref{velocities}) one obtains formally for the total number of
particles in the system
\begin{equation}
  N_{\rm part} = q \int_{0}^{2\pi} d\phi
       \int_{0}^{\pi} d\tau
       \int_{0}^{\infty} dm
       \int_{0}^{2\pi} d\xi
       \int_{0}^{\Psi} dE
       \int_{0}^{\eta_{\rm max}} d\eta \> 
          { {m^2 \sin\tau \fm(E,\eta) \> L_{\rm max}(E)}\over 
            {R \> \mu({\vec r},{\vec v})} } .
\end{equation}
The integrand is independent of the angles $\phi$ and $\xi$, provided
that $\mu$ does not depend on them. For efficient use of the rejection
method one must have phase space coordinates with a finite range, and
an integrand that is as close as possible to constant. This can be
achieved with the coordinate transformations
\begin{eqnarray} 
  &&t = {1\over2} (1-\cos\tau)              ,\qquad
    \eta' = \eta / \sin\tau = \eta m / R    ,\nonumber \\
  &&U = \int_{0}^{m} u(m') \> dm' \> \left / \> 
        \int_{0}^{\infty} u(m') \> dm' \right. 
             ,\qquad u(m) = m^{-1/2} (A+m)^{-1} ,\nonumber \\
  &&V = \int_{0}^{\Psi-E} v(E') \> dE' \> \left / \> 
        \int_{0}^{\Psi} v(E') \> dE' \right.
             ,\qquad v(E) = (B\Psi+\Psi-E)^{-3/2} ,
\end{eqnarray}
which yield 
\begin{equation}
  N_{\rm part} \propto 
       \int_{0}^{2\pi} d\phi
       \int_{0}^{1} dt
       \int_{0}^{1} dU
       \int_{0}^{2\pi} d\xi
       \int_{0}^{1} dV
       \int_{0}^{\eta'_{\rm max}} d\eta' \>
          { {m \> \fm(E,\eta) \> L_{\rm max}(E)}\over
            {\Psi^{1/2} \> u(m) \> v(E) \> \mu({\vec r},{\vec v})} } .
\label{Npart}
\end{equation}
The arbitrary constants $A$ and $B$ can be chosen to optimize the
efficiency of the algorithm. The integrand is a known function of
$(t,U,V,\eta')$, so one may use the rejection method with the
comparison constant
\begin{equation}
  F_{\rm comp} = \max_{\rm phase\>space} \>
          { {m \> \fm(E,\eta) \> L_{\rm max}(E)}\over
            {\Psi^{1/2} \> u(m) \> v(E) \> \mu({\vec r},{\vec v})} } ,
\end{equation}
which is finite for most $\mu({\vec r},{\vec v})$ of interest. Its
value can be found with, e.g., the simplex algorithm of Press \etal
(1992). It can be shown that $\eta'_{\rm max} < 1/q$ for an
axisymmetric system. Particles with an arbitrary mass spectrum
$\mu({\vec r},{\vec v})$ can therefore be generated using the
following algorithm:
\begin{enumerate}
\item Draw random numbers $\cR_1,\cR_2,\cR_3$ and $\cR_4$. Set 
\begin{eqnarray}
  &&\tau = \arccos \> (1-2\cR_1) , \qquad
    m = A \tan^2 (\pi \cR_2 /2)  , \qquad
    \eta' = \cR_3 / q            , \nonumber \\
  &&E = \Psi \left\lbrace (B+1) - B 
           \left[ 1 - \cR_4 \left( 1 - (B/B+1)^{1/2} 
           \right) \right]^{-2} \right\rbrace .
\end{eqnarray}
\item Calculate $\eta_{\rm max}$ from equations~(\ref{mtau}) 
and~(\ref{etamax}). Reject the previous step and return to
step~1 if $\eta' \sin\tau > \eta_{\rm max}$.
\item Draw a random number $\cR_5$. Calculate the integrand in 
equation~(\ref{Npart}). Reject the previous steps and return to
step~1 if the integrand is $< \cR_5 F_{\rm comp}$.
\item Draw random numbers $\cR_6$ and $\cR_7$. Set 
$\phi = 2 \pi \cR_6$ and $\xi = 2 \pi \cR_7$. Calculate the velocities from
equation~(\ref{velocities}).
\end{enumerate}

After the required number of particles ($N$) has been drawn, all
particle masses $\mu_i = \mu({\vec r}_i,{\vec v}_i)$ must be
renormalized such that the total mass $\sum \mu_i$ equals $M$, as
given by equation~(\ref{totmass}). For a system of equal-mass
particles one simply sets $\mu_i = M/N$ for $i=1,\ldots,N$.

\clearpage


\clearpage



\def\figcapone{Histograms of the number of particles as a function of
radius $r$ for $N=\>$512,000 initial conditions derived from the
$f(E,L_z)$ model with $i=55^{\circ}$. Solid curve: equal-mass
particles; dotted curve: particles with a mass spectrum $\mu(E)$ as
described in the text. Vertical dashed lines indicate three relevant
scales: the softening length $\epsilon=0.04''$ of the dark nuclear
object; the scale length $b=0.55''$ within which the mass density has
a central cusp (cf.~eq.~[\refmassdens]); and the outer scale length of
the mass density $c=102''$. The mass spectrum gives the particles at
smaller radii a smaller mass (see Figure~2), thereby increasing the
resolution near the center. Both histograms are normalized to the same
total number of particles.\label{f:histnum}}


\def\figcaptwo{Scatter plot of the particle mass $\mu$ as a function
of radius $r$ for the $i=55^{\circ}$ multi-mass model. The left axis
shows $\mu$ in dimensionless units, the right axis in solar
masses. The vertical dotted lines show the scales $\epsilon$, $b$
and~$c$ as in Figure~1, and also the scales $r_0$ and $r_1$ that enter
into the definition of the mass spectrum
(cf.~eq.~[\refmassspec]). Particles at small radii have smaller masses
than those at large radii.\label{f:scatter}}


\def\figcapthree{The circular velocity (in dimensionless units) of the
$i=55^{\circ}$ model in the equatorial plane, as a function of radius
(in dimensionless units). The dotted vertical line shows the softening
radius $\epsilon$ of the dark object. The short-dashed curve labeled
`$v_{\rm c}(\star)$' is the circular velocity that corresponds to the
mass density given by equation~({\refmassdens}). The solid curve
labeled `$v_{\rm c}({\rm SCF})$' is the numerical approximation
obtained from a SCF expansion with $n_{\rm max}=8$ terms. The
approximation is poor for $\log(r) \lta -1.0$, due to the finite
resolution of the SCF expansion. The long-dashed curve labeled
`$v_{\rm c}({\rm BH})$' shows the circular velocity of the (softened)
black hole. The two (party overlaying) solid curves labeled `$v_{\rm
c}({\rm total})$' correspond to the actual total circular velocity of
the model, and to the approximation obtained with the SCF
expansion. The latter provides a good fit for all $\log(r) \gta -2.7$,
because the circular velocity at small radii is dominated by the
contribution from the dark object. Hence, the numerical representation
of the model is sufficiently accurate for all $r \gta
\epsilon$.\label{f:vcirc}}


\def\figcapfour{Initial and final mass densities for the
$i=55^{\circ}$ model, as a function of mean spheroidal radius $m$
(both in dimensionless units). The initial conditions were drawn from
the mass density given by equation~({\refmassdens}). The inner curves
show the shorter time integration for the multi-mass model, the outer
curve shows the longer time integration for the equal-mass model. The
mass density profile is stable with time.\label{f:dens}}


\def\figcapfive{Axial ratios $q_1$ and $q_2$ of different mass bins for
the $i=55^{\circ}$ model, as a function of time (in dimensionless
units), calculated as described in the text. The range of spheroidal
radius $m$ for each bin is indicated in the panel, as is the fraction
of the total mass in that bin. The mass density~({\refmassdens}) from
which the initial conditions were drawn has $q_1 \equiv 1$ and $q_2
\equiv q = 0.55$.  The axial ratios in the simulation remain stable at
these values.\label{f:axialrat}}


\def\figcapsix{The solid curves show the time evolution of
several representative terms in the SCF decomposition for the
$i=55^{\circ}$ multi-mass model. The mass density is expanded onto the
set of basis functions $\rho_{nlm}$ given by Hernquist \& Ostriker
(1992). The $A_{nlm}$ are the expansion amplitudes; $n$ refers to the
radial part of the expansion, and $l$ and $m$ to the spherical
harmonic part. The amplitudes are shown logarithmically on a base-10
scale, in dimensionless units. Dashed lines in the two left-most
columns (partially overlaid by the solid curves) show the analytically
calculated $m=0$ amplitudes for the mass density given by
equation~({\refmassdens}); the $m=1$ and $m=2$ terms shown in the two
right-most columns are zero for this axisymmetric mass density. The
$A_{800}$ term shows a small secular variation, arising from the
finite resolution of the SCF expansion. The variations of all other
terms are consistent with random fluctuations. In particular there is
no secular variation in the non-axisymmetric ($m \neq 0$)
terms.\label{f:Anlm}}


\def\figcapseven{Initial and final velocity moments (in dimensionless
units) as a function of radius (in dimensionless units), for the
$i=55^{\circ}$ multi-mass model. The left panel shows the mean
azimuthal streaming velocity $\overline{v_{\phi}}$. The right panel
shows the RMS velocities $[\overline{v_{\phi}^2}]^{1/2}$,
$[\overline{v_{\theta}^2}]^{1/2}$ and
$[\overline{v_{r}^2}]^{1/2}$. The latter two are identical in an
$f(E,L_z)$ model. The vertical dotted lines show the softening length
$\epsilon$ of the dark nuclear object. There is some evolution in the
velocities at small radii, due to the small number statistics and the
finite resolution of the SCF expansion. Overall, the velocity
structure of the model is stable.\label{f:kinematics}}


\def\figcapeight{The variation in total kinetic energy ($T$; top),
potential energy ($W$; middle) and the virial ratio (bottom) for the
$i=55^{\circ}$ model, as a function of time (all in dimensionless
units). The displayed virial ratio is not $|2T/W|$, but properly takes
into account the contribution of the external potential of the dark
object as dictated by the virial theorem. The solid curves are for the
multi-mass model, the dotted curves are for the equal-mass model. The
curves for these two cases are not identical, because the particle
distributions statistically sample different radii of the model
(cf.~Figure~1). The virial ratio is constant with time, consistent
with the stability inferred from the other diagnostic
quantities.\label{f:virial}}


\ifsubmode
\figcaption{\figcapone}
\figcaption{\figcaptwo}
\figcaption{\figcapthree}
\figcaption{\figcapfour}
\figcaption{\figcapfive}
\figcaption{\figcapsix}
\figcaption{\figcapseven}
\figcaption{\figcapeight}
\clearpage
\else\printfigtrue\fi

\ifprintfig


\begin{figure}
\epsfxsize=10.0truecm
\centerline{\epsfbox{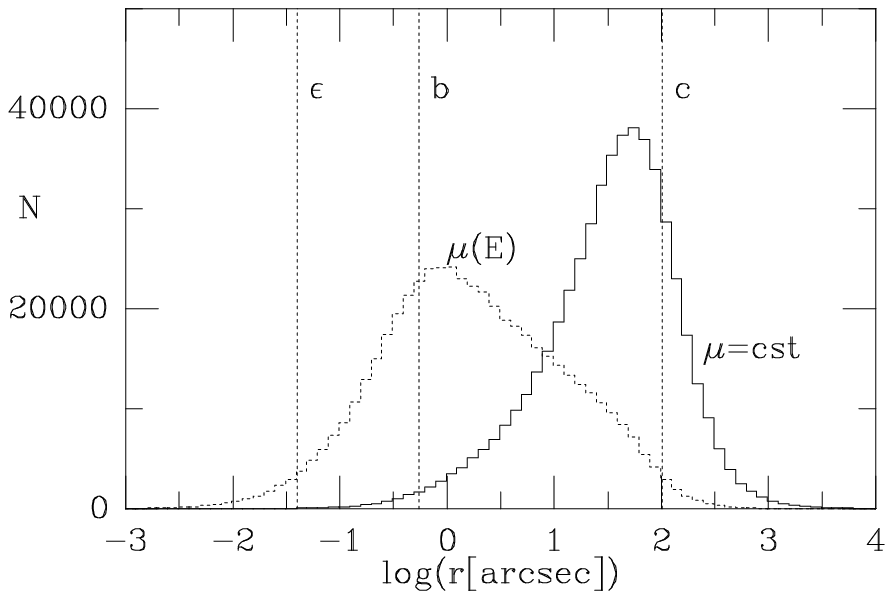}}
\ifsubmode
\vskip3.0truecm
\centerline{Figure~1}\clearpage
\else\figcaption{\figcapone}\fi
\end{figure}


\begin{figure}
\epsfxsize=10.0truecm
\centerline{\epsfbox{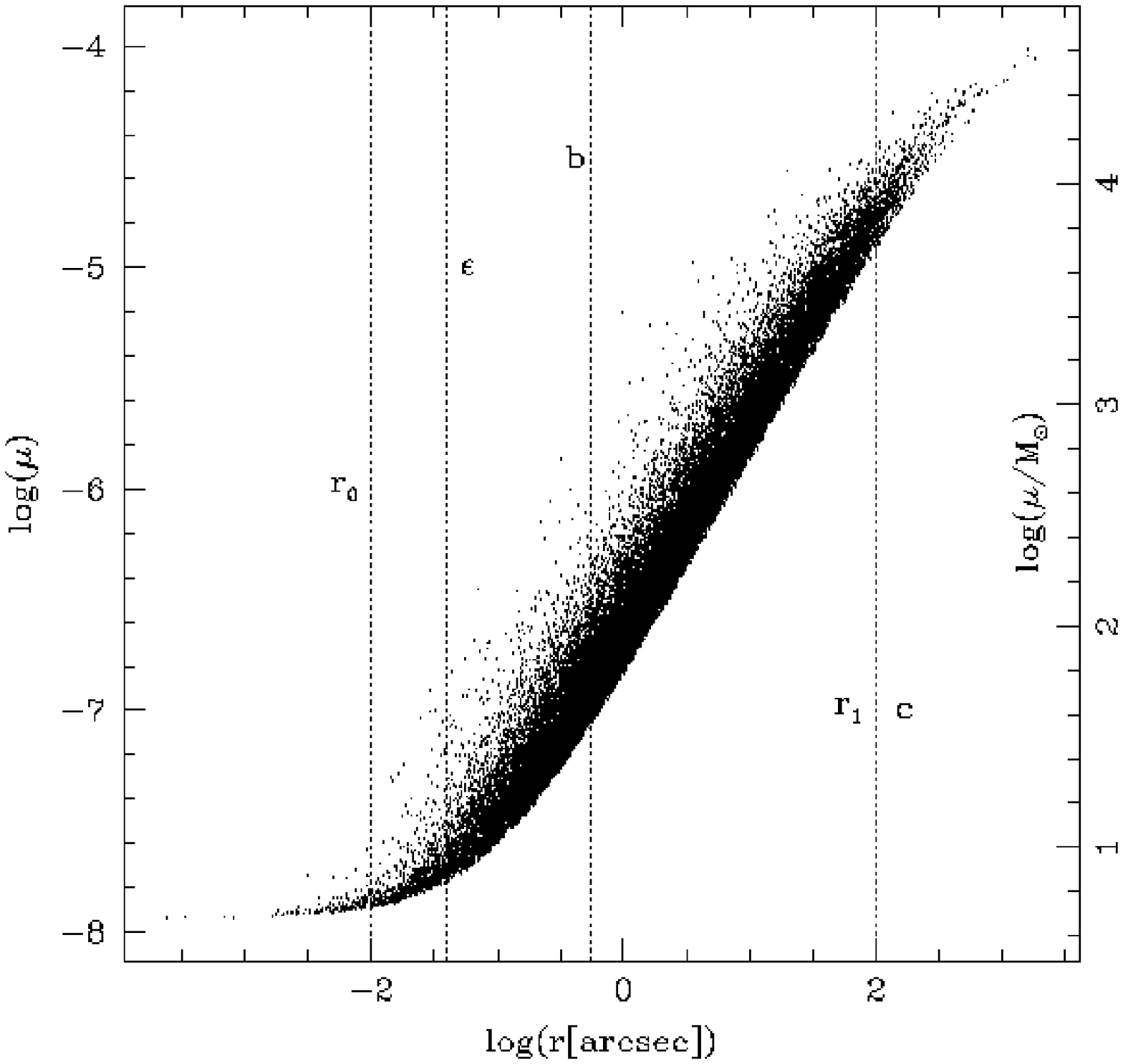}}
\ifsubmode
\vskip3.0truecm
\centerline{Figure~2}\clearpage
\else\figcaption{\figcaptwo}\fi
\end{figure}


\begin{figure}
\epsfxsize=10.0truecm
\centerline{\epsfbox{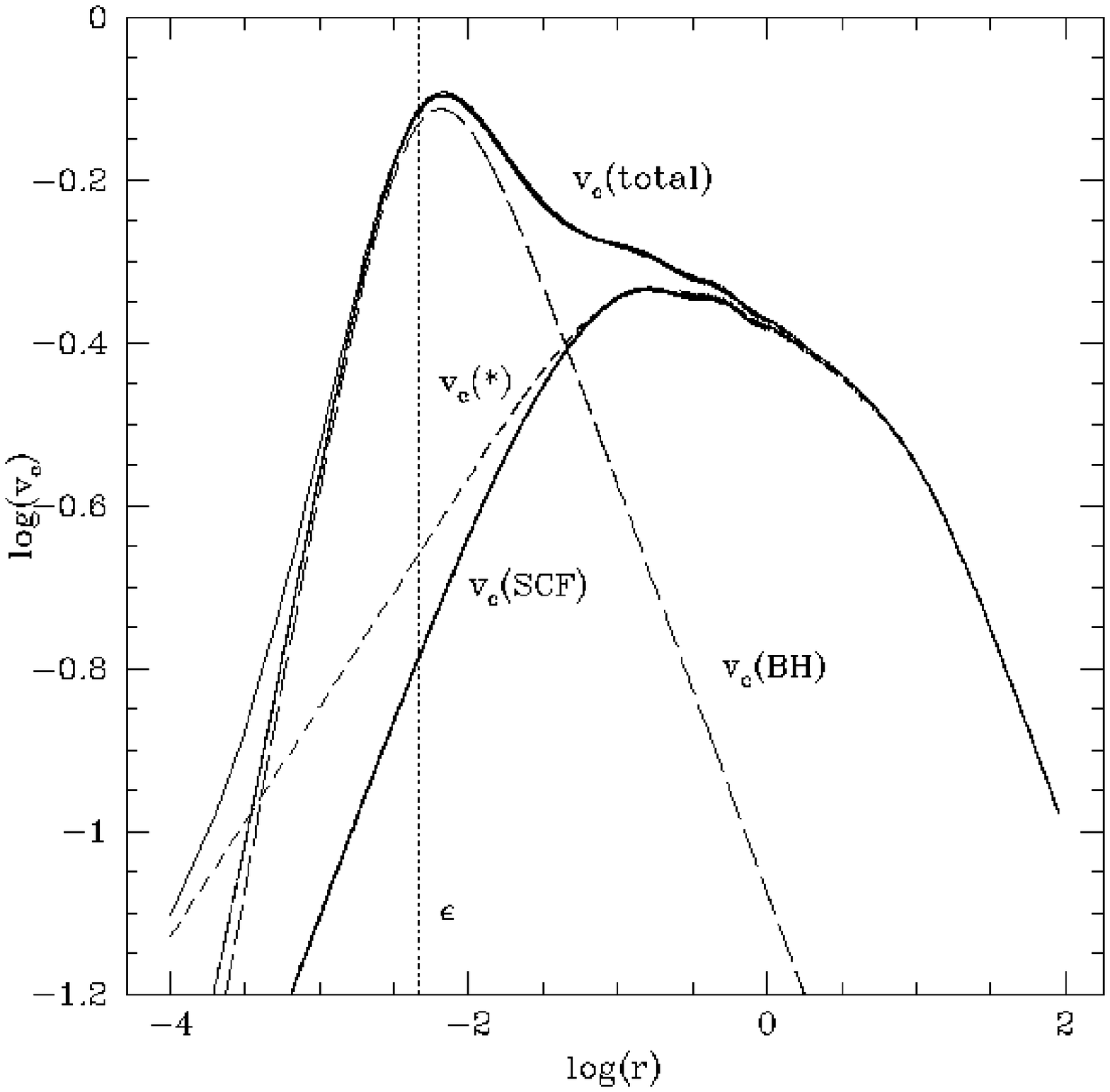}} 
\ifsubmode
\vskip3.0truecm
\centerline{Figure~3}\clearpage
\else\figcaption{\figcapthree}\fi
\end{figure}


\begin{figure}
\epsfxsize=10.0truecm
\centerline{\epsfbox{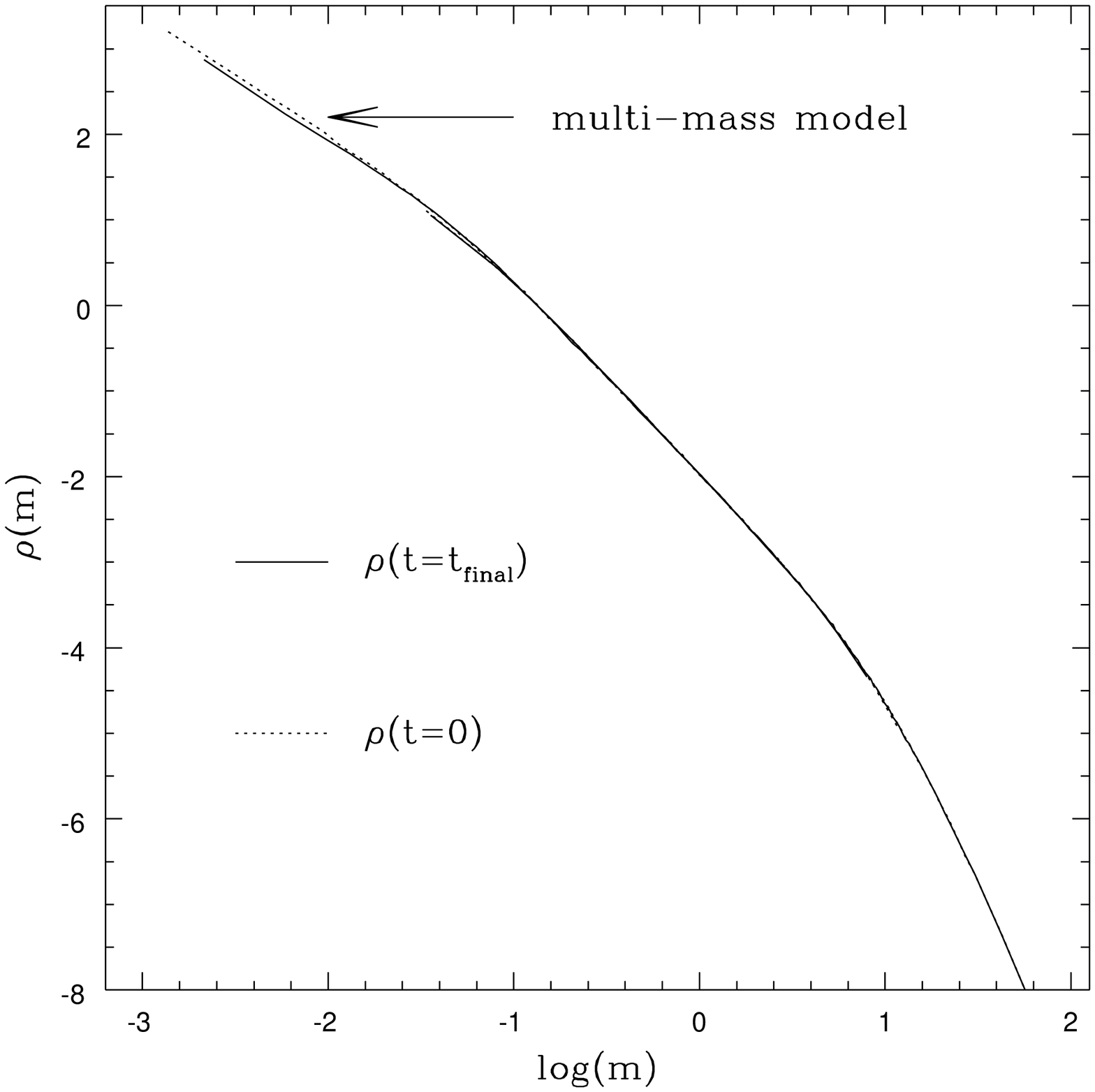}} 
\ifsubmode
\vskip3.0truecm
\centerline{Figure~4}\clearpage
\else\figcaption{\figcapfour}\fi
\end{figure}


\begin{figure}
\epsfxsize=10.0truecm
\centerline{\epsfbox{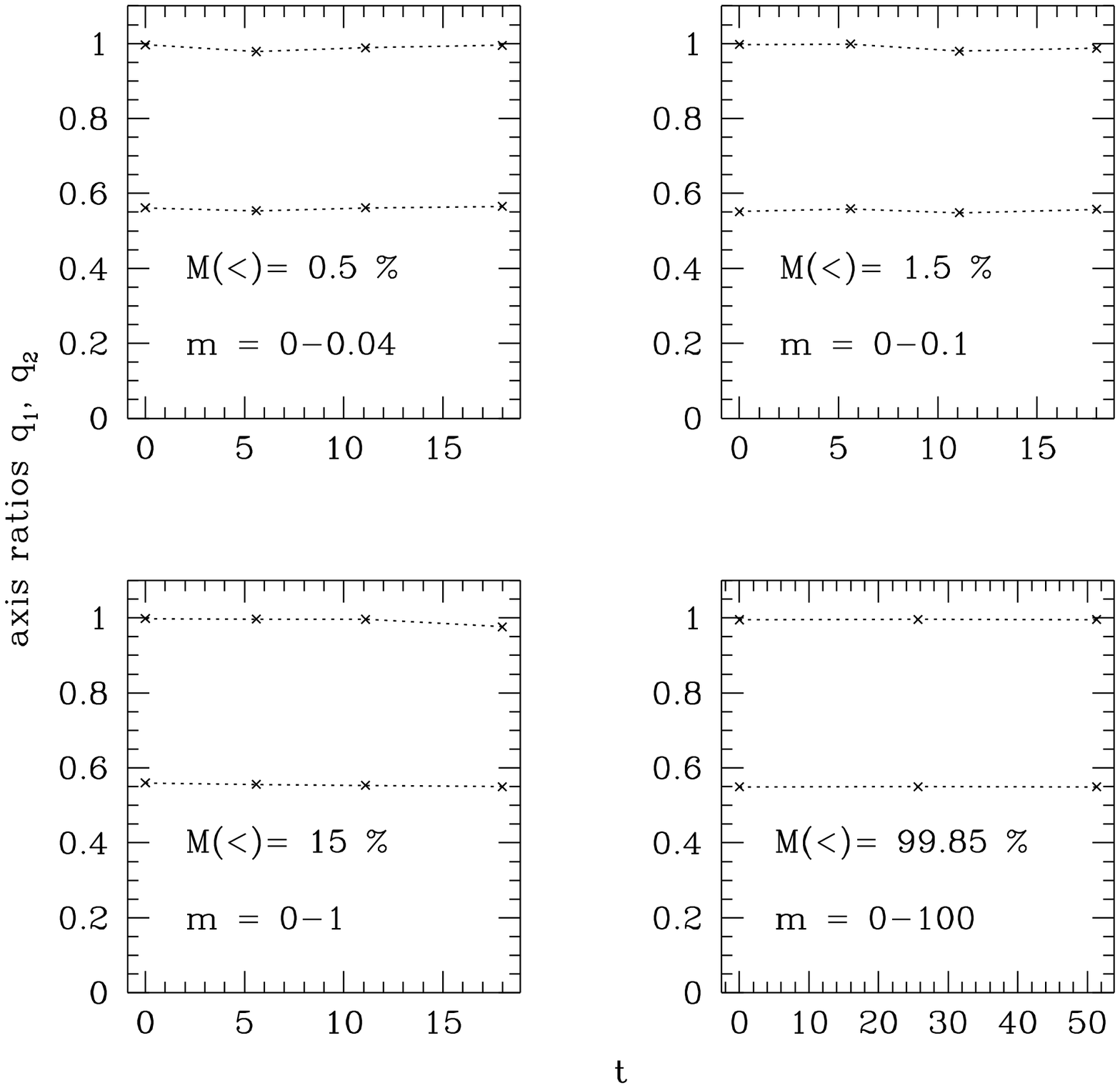}} 
\ifsubmode
\vskip3.0truecm
\centerline{Figure~5}\clearpage
\else\figcaption{\figcapfive}\fi
\end{figure}


\begin{figure}
\epsfxsize=10.0truecm
\centerline{\epsfbox{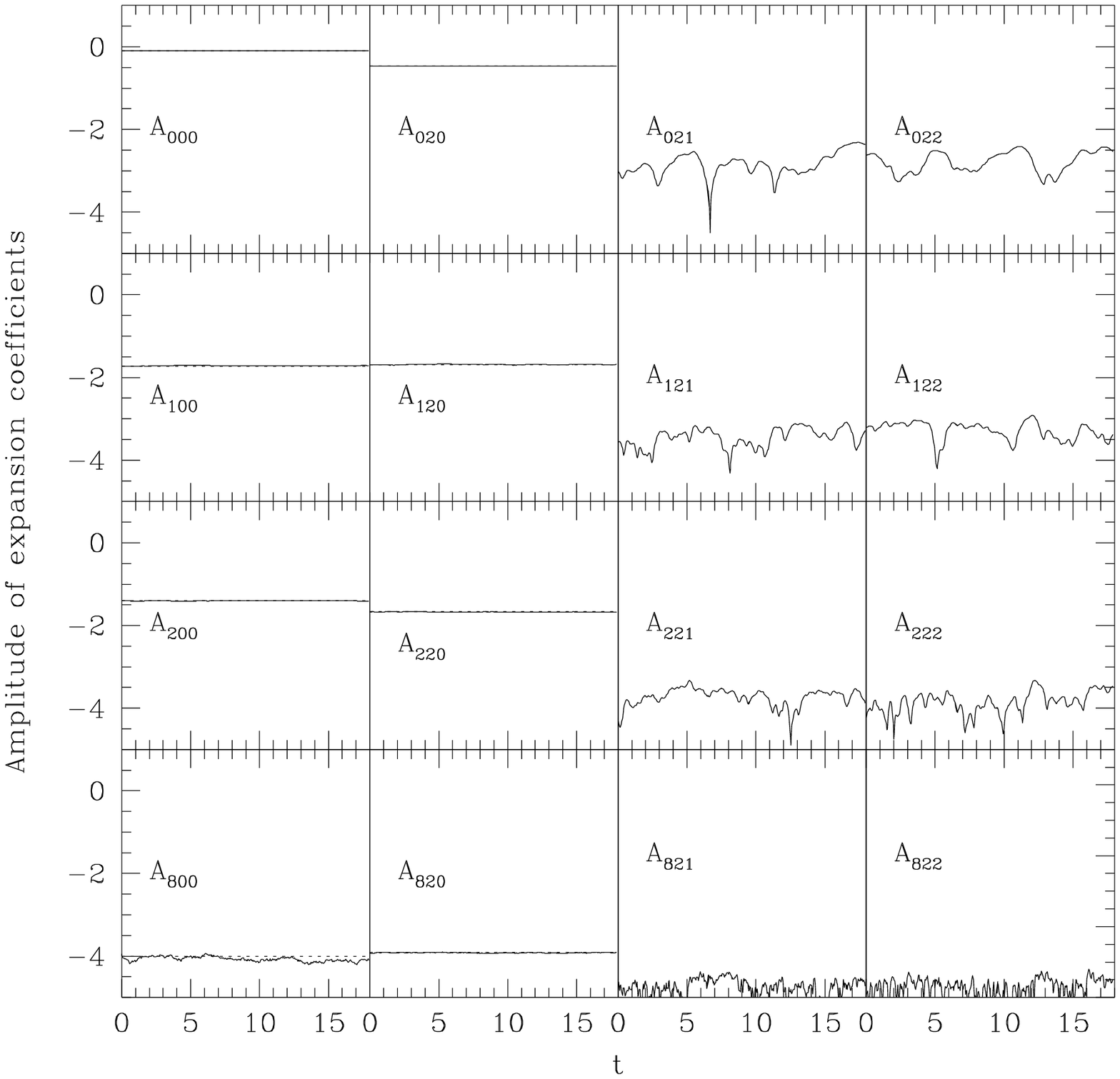}}
\ifsubmode
\vskip3.0truecm
\centerline{Figure~6}\clearpage
\else\figcaption{\figcapsix}\fi
\end{figure}


\begin{figure}
\epsfxsize=10.0truecm
\centerline{\epsfbox{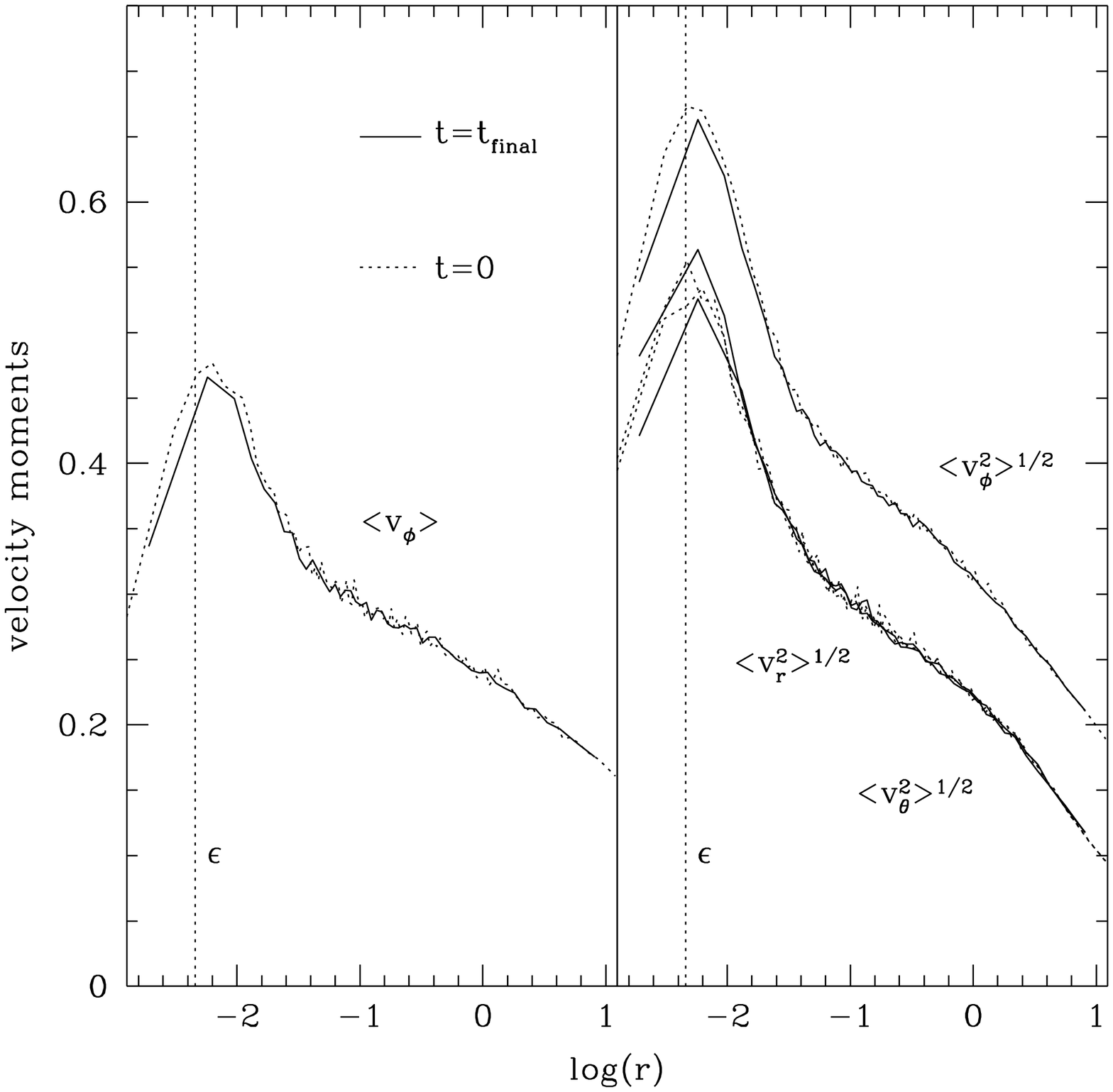}}
\ifsubmode
\vskip3.0truecm
\centerline{Figure~7}\clearpage
\else\figcaption{\figcapseven}\fi
\end{figure}


\begin{figure}
\epsfxsize=10.0truecm
\centerline{\epsfbox{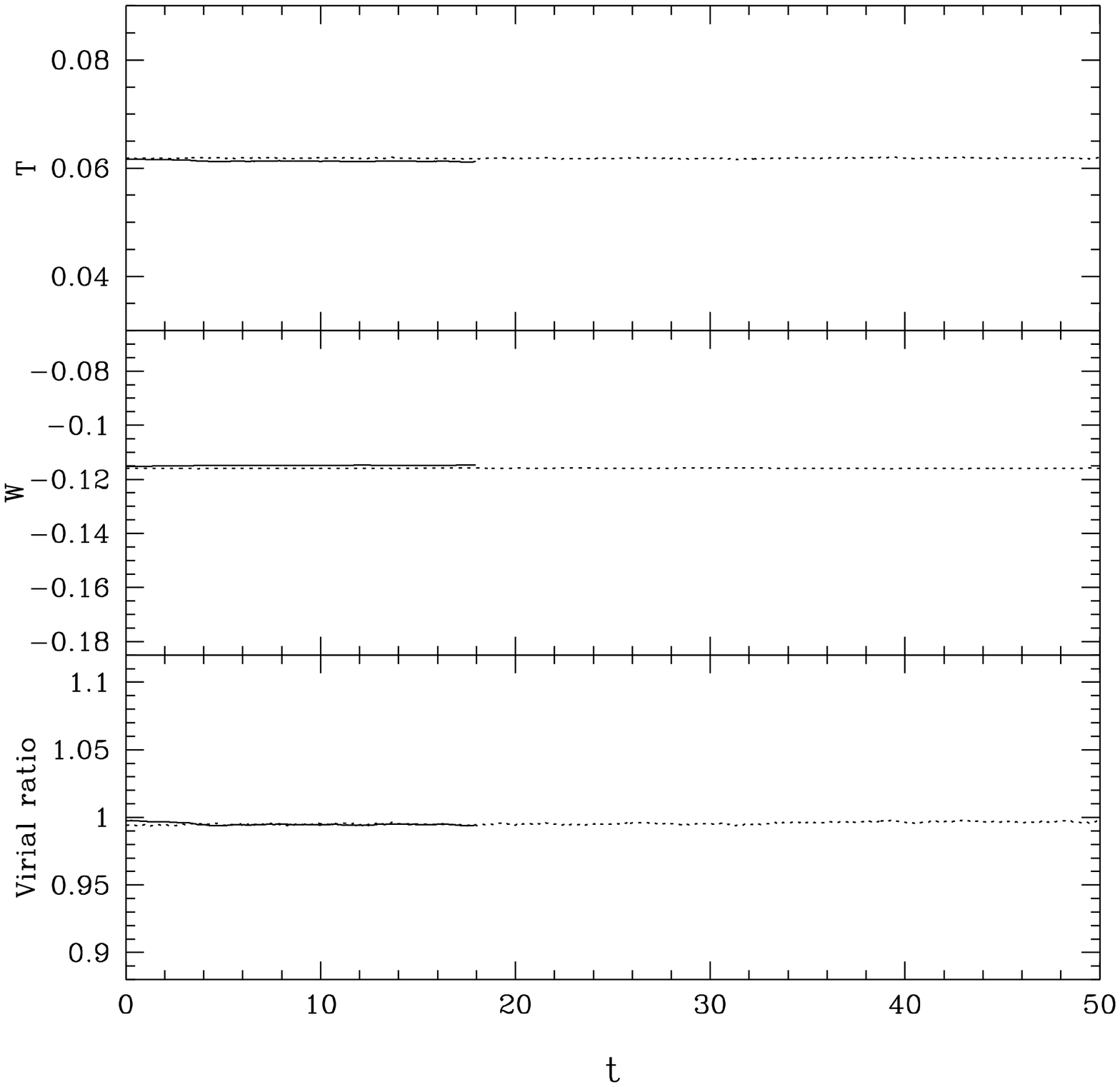}}
\ifsubmode
\vskip3.0truecm
\centerline{Figure~8}\clearpage
\else\figcaption{\figcapeight}\fi
\end{figure}


\fi

\end{document}